\documentclass[multphys,vecphys]{svmult}

\usepackage{makeidx}     
\usepackage{graphicx}    
\usepackage{multicol}    

\makeindex             

\begin{document}

\title*{350~$\mu$m Galactic Center Dust Observations with SHARC~II}
\author{Johannes Staguhn \inst{1}\and Dominic Benford\inst{1} \and Mark Morris\inst{2}\and
Keven Uchida\inst{3}}
\institute{NASA/GSFC, Code 685, Greenbelt, MD 20771, USA
\texttt{staguhn@stars.gsfc.nasa.gov}
\and Division of Astronomy \& Astrophysics, UCLA, Los Angeles, CA 90095-1562, USA
\and Dept. of Astronomy, Cornell University, Ithaca, NY 14853-6801, USA}
%
%
\maketitle

We present first and preliminary submillimeter continuum images of the Galactic Center region obtained with the new Caltech Submillimeter Observatory facility camera SHARC~II.  The instrument allows 350~$\mu$m observations with unprecedented sensitivity and instantaneous spatial coverage. 
The stability of the SHARC~II detectors combined with the large number of available pixels allows a non-differential scanning mode that does not require the observation of a reference off-source position. 
Here we present large-scale 350~$\mu$m dust continuum images from the Sgr~A  and Sgr~C regions, as well as the
detection of sub-mm dust continuum emission from the  IRAS 25$\mu$m source AFGL5376. This source is produced in a large-scale shock that extends well above the Galactic plane.

\section{Introduction}
\label{sec:1}
Studies of the interstellar medium in the nuclei of Galaxies can be used
to derive  physical parameters for galaxy
evolutionary models.  The presence of
stronger magnetic fields than in the Galactic disk and strong shear due
to differential rotation allow efficient studies of their effects on
the interstellar medium and on star formation activity. 
Observations of the submillimeter continuum emission with SHARC~II \cite{Darren_SHARCII} play an important role in the determination of dust temperatures and opacity indices in Galactic Center molecular clouds. The heating of the dust is typically dominated by the
interstellar UV field, and hence is attributable to relatively recent star formation.  Other heating mechanisms such as interaction of molecular clouds with large-scale shocks are not well explored. We therefore targeted a large-scale shock region with no apparent star formation activity. Furthermore, submillimeter dust continuum data can be used to identify target positions for polarimetric follow-up observations aimed at  obtaining  magnetic field directions in molecular clouds \cite{Novak}. 
\section{Sgr A  and Sgr C}
\label{sec:2}
\begin{figure}
\centering
\includegraphics[width=10cm]{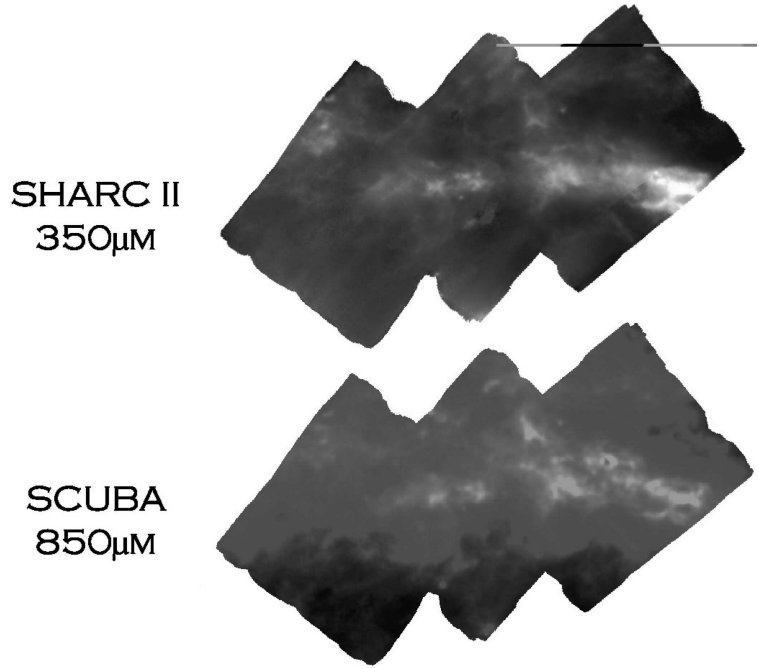}
%
%
\caption{{\it top:} SHARC~II  350~$\mu$m image of the Sgr region. 
             {\it bottom:} SCUBA 850~$\mu$m image  of the same region (from \cite{Pierce-Price}). }
\label{fig:1}       
\end{figure}
\begin{figure}
\centering
\includegraphics[width=11.4cm]{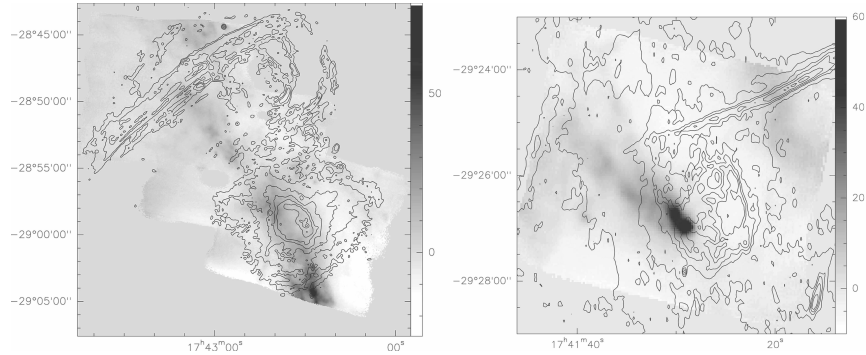}
%
\caption{{\it left:} SHARC~II  350~$\mu$m image of the Sgr A region (grey scale) with the 20 cm radio image from \cite{Yusef-Zadeh_galcenarc} superimposed in contours. The most prominent features in the  radio image are Sgr~A, the Galactic Center Thermal Arches, and the Nonthermal Radio Arc. {\it right:} SHARC II 350~$\mu$m observations of the region around Sgr~C (grey scale). Superimposed is the 1.5 GHz radio image from \cite{Liszt_Spiker}. Note the anti-correlation between the radio emission and the dust continuum towards the NW of the Sgr C HII region.}
\label{fig:2}       
\end{figure}
We observed an area of about $20' \, \times \, 20'$ at 350 ~$\mu$m which covers the main dust features around Sgr~A.
Figure \ref{fig:1} shows the SHARC~II  image and, for comparison, the corresponding area as observed by SCUBA at 850~$\mu$m \cite{Pierce-Price}. Figure \ref{fig:2} (left)  shows the SHARC~II image with an overlay of the 20 cm image from \cite{Yusef-Zadeh_galcenarc}.
Figure \ref{fig:2} (right) shows the region around Sgr~C at 350~$\mu$m. Superimposed on the image are the 1.5~GHz radio contours from \cite{Liszt_Spiker}  which show the Sgr~C HII region and the associated nonthermal filaments. An anti-correlation of the dust emission and radio emission is evident north of the HII region.
%
%
%
%
\section{AFGL5376}
\label{sec:3}
\begin{figure}
\centering
\includegraphics[width=5.8cm]{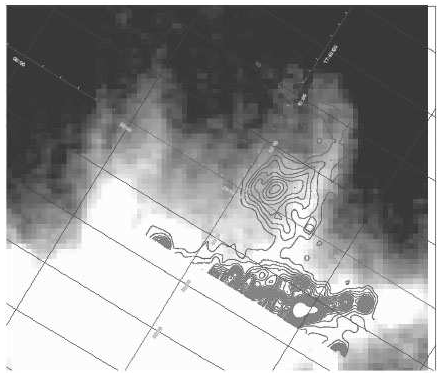}
%
%
\caption{Large scale Nobeyama 10 GHz image \cite{Nob_10GHz} (grey scale) of 
the Galactic center with IRAS 25 $\mu$m contours overlayed. The Galactic Center Lobes East and West can be seen as vertical structures in the grey-scale 
representation of the Nobeyama data. AFGL5376 is the prominent 25 
$\mu$m feature at center right in the image, to the immediate East of the radio 
emission from the Galactic Center Lobe West.  The Galactic plane is 
horizontal, and is indicated by the highest 25~$\mu$m contours.}
\label{fig:3}       
\end{figure}
\begin{figure}
\centering
\includegraphics[width=11.4cm]{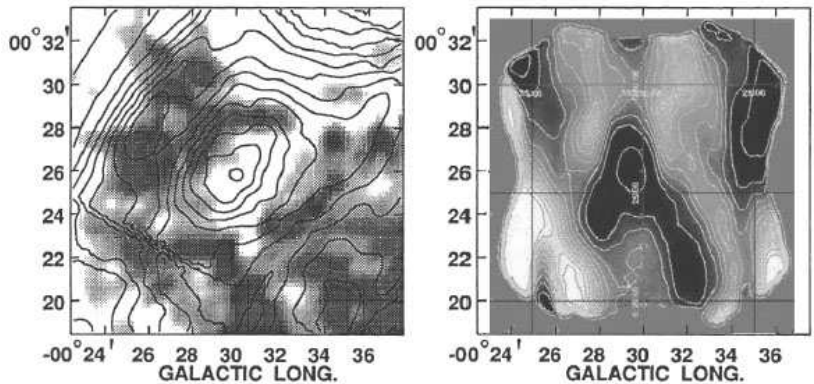}
%
\caption{{\it left:} AFGL5376  in  $^{12}$CO(2-1) emission (grey scale, from \cite{Uchida_AFGL5376}) with an overlay of the  IRAS 25~$\mu$m emission (contours) {\it right:} SHARC~II 350~$\mu$m continuum image of the region (grey scale and contours). Note that the grey scale representation of the right image shows emission light, whereas it is dark in the left image.}
\label{fig:4}       
\end{figure}
The IRAS source AFGL5376 is probably the most prominent 
concentration of warm dust near the Galactic Center which clearly shows 
no evidence for nearby star formation.  This curious structure rises almost a full degree 
(150 pc) perpendicular to the Galactic plane  (Figure \ref{fig:3}).  Its strong  25 $\mu$m emission 
indicates the presence of warm ($T\sim$100 K)  dust which is clearly 
associated with high velocity $^{12}$CO(2-1) emission (Figure \ref{fig:4} left, 
from \cite{Uchida_AFGL5376}). The  molecular line emission traces gas that 
surrounds the warm dust emission peak.  Uchida et al. \cite{Uchida_AFGL5376} show that 
the kinematics of the gas as well as the position of AFGL5376 provide 
strong evidence that it is associated with a large scale shock that 
coincides with the western edge of the radio continuum Galactic Center 
Lobe. However, the mass of the warm dust in AFGL5376 is only $10^{-2}$ M$_{\odot}$, and until 
recently it was unclear whether larger amounts of colder dust are 
associated with this source. In August 2003 we obtained a snapshot of this region at 350~$\mu$m with SHARC~II in order to address this question. The observations (Fig. \ref{fig:4}, right) show strong  350~$\mu$m continuum emission which, like the molecular 
line emission,  surrounds the 25 $\mu$m peak. A void in the emission 
stretching perpendicular to the Galactic plane is clearly visible in 
the SHARC~II image. Its position coincides with one of the two shock 
fronts which are described in \cite{Uchida_AFGL5376}. A determination of 
the spatial extent, column densities and temperatures of the dust and the
nature of its association with the Galactic Center Lobe and the associated large 
scale shock is under way.  This will enable us to study the energetics involved 
in the shock generation and will be important for understanding the 
relationship between the kinematics and magnetic fields in the Galactic 
center region.
 \bibliographystyle{}
 \bibliography{}

\begin{thebibliography}{99.}
%
%
%

\bibitem{Darren_SHARCII} C.D. Dowell,  et al.,  SPIE,   \textbf{4855}, 73 (2003)

\bibitem{Nob_10GHz} T. Handa,  Y. Sofue, N. Nakai, H. Hirabayashi, M.  Inoue,  PASJ,  \textbf{39}, 709 (1987)

\bibitem{Liszt_Spiker} H. S. Liszt, R.W. Spiker, ApJS, \textbf{98}, 259 (1995) 

\bibitem{Novak} G. Novak, et al., ApJ, \textbf{583L}, 83 (2003)

\bibitem{Pierce-Price}  D. Pierce-Price et al.,  \textbf{545L}, 121 (2000)

\bibitem{Uchida_AFGL5376} K.I. Uchida, M.R.  Morris, E. Serabyn, 
J.  Bally, APJ,  \textbf{421}, 505 (1994)

\bibitem{Yusef-Zadeh_galcenarc} F. Yusef-Zadeh, M. Morris,  D. Chance, Nature, \textbf{310}, 557 (1984) 


\end{thebibliography}
%

\printindex
\end{document}